\documentclass[preprint,12pt]{elsarticle}

\usepackage{amsmath,amssymb,amsfonts}
\usepackage{graphicx}
\usepackage{placeins}
\usepackage{booktabs}
\usepackage{siunitx}
\usepackage{subcaption}
\usepackage{algorithm}
\usepackage{algpseudocode}
\usepackage{xcolor}
\usepackage{hyperref}
\usepackage{lineno}
\modulolinenumbers[1]

\newcommand{\bsig}{\boldsymbol{\sigma}}
\newcommand{\bx}{\mathbf{x}}
\newcommand{\bX}{\mathbf{X}}
\newcommand{\bu}{\mathbf{u}}
\newcommand{\br}{\mathbf{r}}
\newcommand{\bF}{\mathbf{F}}
\newcommand{\bI}{\mathbf{I}}
\newcommand{\bK}{\mathbf{K}}

\newcommand{\bC}{\mathbf{C}}

\newcommand{\dofs}{n_{\mathrm{dof}}}
\newcommand{\Nlinks}{N}

\journal{Journal of the Mechanics and Physics of Solids}

\begin{document}

\begin{frontmatter}

\title{Matrix-free GPU energy minimization enables direct finite-strain simulation of the chain network underlying the Arruda-Boyce model}

\author[a]{Yizhi Chen\fnref{equal}}
\author[b]{Yuxuan Huang\fnref{equal}}
\author[c]{Zihao Chen}
\author[c]{Zhili Xiao}
\author[b]{Naiyan Wu}
\author[a]{Dushyanth Rudra Mahesh Velugubantla}
\author[b]{Xiangjun Peng}
\author[a]{Ellen Arruda}
\author[b]{Guy M.\ Genin\corref{cor1}}
\ead{genin@wustl.edu}

\address[a]{Department of Mechanical Engineering, University of Michigan, Ann Arbor, MI, USA}
\address[b]{NSF Science and Technology Center for Engineering Mechanobiology and Department of Mechanical Engineering \& Materials Science, Washington University, St. Louis, MO, USA}
\address[c]{Department of Electrical and System Engineering, Washington University, St. Louis, MO, USA}

\cortext[cor1]{Corresponding author}
\fntext[equal]{These authors contributed equally to this work.}

\begin{abstract}
The Arruda-Boyce model derives the elasticity of rubber from a representative cell of eight non-Gaussian chains, assigning all eight a single stretch fixed by the first invariant of the deformation.
This first-invariant closure makes the model isotropic and analytically closed, but it is an assumption about how the underlying chain network deforms, and it has never been tested directly against that network, in part because the network could not be solved at realistic scale.
Here, we develop a fast and robust method for the mechanics of large, nonlinear networked solids, and apply it to this problem.
The body is tiled by eight-chain cells whose inverse-Langevin chains are evaluated from their individual end-to-end lengths, near-incompressibility is enforced by a per-cell volumetric penalty, and equilibrium is obtained by minimizing the total stored energy with a
matrix-free, GPU-resident L-BFGS solver that never forms a global tangent.
Because computational cost is governed by local gradient evaluations and not by a global linear solve, explicit-network simulation at $10^{7}$ degrees of freedom becomes routine on a single GPU.
Against Abaqus solving the closed-form continuum model, in which the finite-element code does far less work per material point, a coarse-to-fine cascade carries the method from parity to a $65$-fold advantage at $7\times10^{6}$ displacement degrees of freedom on the hardware tested.
Requiring only a gradient, the minimization also stays well posed on the rank-deficient tangents that defeat Newton iteration, returning an admissible configuration for a pin-jointed auxetic lattice at every size tested, where implicit Newton succeeds only for the two smallest.
Solving the network revealed that, under homogeneous deformation aligned with the cell axes, the eight chains share a common stretch and the Arruda-Boyce response is recovered, while under the strongly inhomogeneous loading of a perforated plate, where non-affine kinematics can emerge, the network and the closed-form model differ by only a few percent.
The Arruda-Boyce first-invariant collapse, adopted originally on physical grounds, is thereby supported by direct computation; the small residual difference, which includes discretization error, indicates that the closure discards little.
The numerical method extends to anisotropic, disordered architectures for which no closed-form closure exists and affine kinematics is least justified.
\end{abstract}


\end{frontmatter}

\section{Introduction}
\label{sec:intro}

The mechanics of rubber-like and fibrous solids is, at the microstructural level, the mechanics of a network. Vulcanized rubber is a dense three-dimensional (3D) network of crosslinked polymer chains whose statistical configurations carry the free energy of deformation~\cite{treloar1975physics};
biological tissues, gels, and many architected materials are likewise built from discrete load-bearing elements~\cite{picu2011}. In computation, the network is often replaced by a continuum constitutive law derived from the network, and the discreteness, together with the non-affine kinematics it permits, is integrated away. This replacement makes large-scale simulation tractable, and has been highly effective for polymers \cite{arrudaboyce1993}. However, in some cases it is simply an approximation whose consequences have been difficult to examine directly.

The Arruda-Boyce eight-chain model is the canonical instance of this replacement~\cite{arrudaboyce1993}. Its representative volume element is a cube spanned by eight identical chains running along the body diagonals from the corners to the center, each obeying non-Gaussian Langevin statistics~\cite{treloar1975physics}. Two assumptions are used to convert this discrete cell into a continuum law. The deformation is taken to be affine, so that each chain stretches with the macroscopic deformation gradient along its own direction; the eight resulting stretches are then replaced by a single representative value, their orientation average over the cell. For the body-diagonal directions $\mathbf{n}_k=(\pm1,\pm1,\pm1)/\sqrt{3}$ this average is
\begin{equation}
\frac{1}{8}\sum_{k=1}^{8}\mathbf{n}_k^{\top}\bC\,\mathbf{n}_k
=\frac{1}{3}\,\mathrm{tr}\,\bC=\frac{I_1}{3},
\label{eq:i1collapse_intro}
\end{equation}
so that the representative chain stretch is $\lambda_{\mathrm{chain}}=\sqrt{I_1/3}$ and the stored energy depends on the deformation only through the first invariant $I_1=\mathrm{tr}\,\bC$. The off-diagonal components of $\bC$, which carry the orientation of the loading relative to the cell axes, cancel identically in the sum. This first-invariant collapse renders the continuum energy isotropic: the cubic geometry of the cell is erased from the macroscopic response. We asked what the consequences of removing this assumption are.

That this affine, orientation-averaged kinematics is an idealization has long been recognized, but while the literature contains refined averaging schemes, the network has not yet been solved directly. The Wu-van der Giessen full-network model~\cite{wuvandergiessen1993} replaces the eight discrete chains with a continuous distribution of chain orientations over the unit sphere. The non-affine micro-sphere model of Miehe et al. has an adjustable, non-affine stretch and a tube constraint that interpolate between the limits of affine limit and fully non-affine deformation~\cite{miehe2004}. Tkachuk and Linder derived the micro-to-macro kinematic link from a maximal-advance-path constraint~\cite{tkachuk2012}. Each of these replaces the Arruda-Boyce closed-form averaging scheme with a more elaborate one, but none solves the underlying network directly. Explicit random fiber networks have been solved directly, revealing strongly non-affine deformation and the controlling role of network connectivity~\cite{picu2011}, but that work targets disordered networks, for which no closed-form continuum limit exists against which the discrete solution could be checked. The eight-chain cell is the complementary case: the one network topology whose homogenized closure is known exactly. This makes a direct test possible.

Testing this requires solving the explicit network, and at finite strain and realistic size this has been the obstacle. The finite-element method remains the standard route to nonlinear boundary-value problems in finite elasticity~\cite{bonet2008nonlinear,wriggers2008nonlinear}, but in its implicit form equilibrium is reached by Newton iteration, and each load increment requires residual evaluation, assembly of a global tangent stiffness, and solution or factorization of a large sparse system.  For a near-incompressible chain network this structure is doubly taxing on computational resources: incompressibility degrades conditioning and element behavior, the inverse-Langevin stiffening sharpens it further as chains approach their locking stretch, and the global solve dominates cost and memory as the model grows to millions of degrees of freedom. Discrete and lattice methods express the energy as a sum of local interactions and are natural for parallel hardware~\cite{cundall1979,silling2000,ostoja2002lattice,buxton2001lattice}, but they have been used mainly for fracture, heterogeneity, and granular or microstructural mechanics, and are typically still advanced by Newton-type solves rather than serving as scalable solvers for continuum-scale hyperelastic problems. Recent GPU and differentiable finite-element frameworks, including JAX-based implementations~\cite{xue2023jaxfem}, accelerate the workflow with automatic differentiation and accelerator hardware but retain the finite-element discretization and its global residual or tangent-solve structure.

For a hyperelastic material there is a way to circumvent the global solution. Equilibrium is a stationary point of the total potential energy, so if potential energy can be written as a sum of local contributions whose gradient is cheap to evaluate in parallel, equilibrium can be found by minimizing the energy directly rather than by repeatedly assembling and factorizing a global tangent. The discrete network supplies this structure: its energy is a sum of 1D chain terms and per-cell volumetric terms, each depending on only a handful of nodes. Limited-memory quasi-Newton methods are promising because they build a search direction from gradients and a short history of vector pairs without ever forming or factorizing a Hessian~\cite{liu1989}. The strategy of relaxing a system by minimizing its local-interaction energy is mature: atomistic statics has long done so through conjugate-gradient relaxation and fast inertial minimization such as FIRE~\cite{bitzek2006}, and the quasicontinuum method applies the same idea to the continuum scale~\cite{tadmor1996}. In computer graphics, where physical accuracy is not the objective, mass-spring and projective-dynamics solvers minimize closely related energies on the GPU for real-time simulation~\cite{bouaziz2014}. However, this machinery has not yet been applied as a quantitative, scalable boundary-value-problem solver for a Langevin chain network.

Here we develop this idea into a general GPU-accelerated solver for finite deformation of chain networks and apply it to the eight-chain cell solved without the first-invariant closure. The body is constructed from eight-chain cells; each chain is an inverse-Langevin spring evaluated from its own current end-to-end length, not from a cell-averaged invariant, so that under inhomogeneous deformation the eight chains of a cell may carry distinct stretches. A per-cell volumetric penalty enforces near-incompressibility. The total energy is minimized by a matrix-free L-BFGS solver implemented in JAX~\cite{jaxopt}. Here, the dominant operations are vectorized local energy and gradient evaluations rather than global assembly and factorization; an optional coarse-to-fine cascade, with the equilibrium field from a coarse mesh initializing the next finer one, accelerates convergence at large problem sizes. After convergence, a cell-wise Cauchy stress is recovered from the converged chain forces by virial homogenization~\cite{irvingkirkwood1950}. The method is benchmarked on a perforated plate under finite uniaxial stretch, a geometry that combines large deformation, near-incompressibility, a curved traction-free boundary, and a stress concentration.

Two findings follow. First, because the method trades the global tangent solver for local gradient evaluations, explicit chain-network simulation at $10^{7}$ degrees of freedom becomes routine on a single GPU; at the largest sizes compared, the solver is more than an order of magnitude faster than Abaqus/Implicit solving the closed-form continuum model. Because energy minimization requires only a gradient, it also remains well posed on the rank-deficient tangents that cause Newton iteration to fail; we demonstrate this on a pin-jointed auxetic lattice. Second, when the deformation is homogeneous and aligned with the cell axes, the eight chains attain the common stretch $\sqrt{I_1/3}$ and the method reproduces the Arruda-Boyce response. Under the inhomogeneous loading of the perforated plate, where the curved boundary and the stress concentration permit non-affine kinematics, the network differs from the continuum Arruda-Boyce model by about 3 to 4\% (3.5\% in peak stress concentration factor, 2.6\% in median field difference), a figure that includes discretization error at the stair-stepped boundary. The result supports the Arruda-Boyce first-invariant collapse in this setting.

\section{Discrete spring-lattice model}
\label{sec:model}

\subsection{Eight-chain unit cell}
\label{sec:8chain}

Arruda and Boyce idealize a crosslinked network as a representative cubic cell of eight identical chains along the body diagonals~\cite{arrudaboyce1993}. In their continuum model, this cell is analyzed to produce a closed-form law: given a deformation gradient $\bF$, the eight chains are assigned the common stretch $\sqrt{I_1/3}$, and Cauchy stress is evaluated analytically from non-Gaussian Langevin statistics; this stress enters a finite-element calculation. The cell itself never appears; only the stretch and stress distilled from it do.

\begin{figure}[htbp]
\centering
\includegraphics[width=\linewidth]{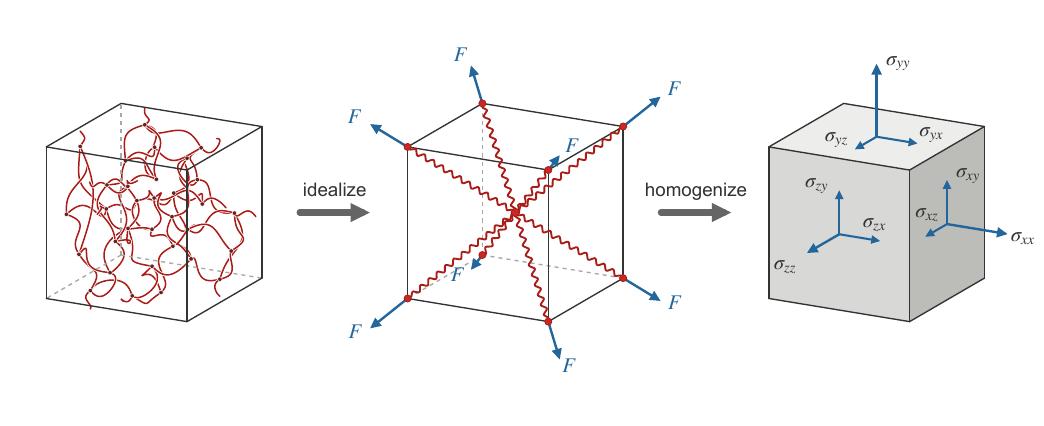}
\caption{
Discrete-network representation of rubber.
\textbf{Left:} a crosslinked polymer network.
\textbf{Center:} a representative eight-chain cell, with chains connecting
the body center to the corners; outward forces $F$ balance chain tension.
\textbf{Right:} the continuum Cauchy stress obtained by homogenizing
the chain forces.
}
\label{fig:model}
\end{figure}

Here, we discard this distillation and retain the cell, keeping the eight chains as discrete, 1D entropic springs. The domain is built by stacking these cells (Fig.~\ref{fig:model}). The kinematic unknowns are the current nodal positions, obtained by direct minimization of total stored energy. Macroscopic stress is recovered by homogenizing converged chain forces. Unlike the continuum model, force in each chain is evaluated from its own current end-to-end length rather than from a cell-averaged invariant. Under homogeneous deformation aligned with the cell axes, the eight chains share a common stretch and the construction reduces to the classical eight-chain kinematics; under a general deformation they need not, and the chains of a single cell may carry distinct stretches and forces.

\subsection{Eight-chain lattice construction}
\label{sec:mesh}

The reference configuration $\Omega_0\subset\mathbb{R}^3$ is tiled by cubic cells of edge $h$ on a regular hexahedral grid of $N_X\times N_Y\times N_Z$ cells. A cell is retained when its centroid lies within the material; cells falling inside a void, such as the central hole of the perforated-plate benchmark to be studied, are dropped from the connectivity, so a curved boundary is resolved as stair-stepped. The set of retained cells is $\mathcal{E}$ (number of cells $|\mathcal{E}|=N_e$) and the nodes $\mathcal{V}=\{\bX_i\}$ are the $N_c$ grid corners, shared between neighbouring cells, together with one body-center node at the centroid of each retained cell, so $N_{\mathrm{node}}=N_c+N_e$ and the problem size is $\dofs=3N_{\mathrm{node}}$. The kinematic unknowns are the current nodal positions $\bx_i$, or, equivalently, the displacements $\bu_i=\bx_i-\bX_i$.

Each retained cell contributes eight chains, connecting its body-center node to its eight corners; the chain set is $\mathcal{C}$ with $|\mathcal{C}|=8N_e$. The reference length $\ell_0^{(c)}$ of chain $c$ is the distance between its two endpoints in $\Omega_0$; for a cubic cell all eight chains have $\ell_0$ equal to half the body-diagonal. Corner nodes are shared while body-center nodes and chains are not, so for a fully dense grid $N_c\approx N_e$, $\dofs\approx 6N_e$, and $|\mathcal{C}|\approx 8N_e$.

\subsection{Chain stored energy}
\label{sec:chainenergy}

Each chain $c$ is treated as a non-Gaussian entropic spring obeying the eight-chain Arruda-Boyce energy. The limiting chain length:
\begin{equation}
r_{c}^{\max}
=
\ell_{0}^{(c)}\sqrt{\Nlinks},
\label{eq:rmax}
\end{equation}
where $\Nlinks$ is the number of statistical (Kuhn) links per chain. The
normalized chain extension is then
\begin{equation}
\lambda_{c}
\;\equiv\;
\frac{r_{c}}{r_{c}^{\max}}
=
\frac{r_{c}}{\ell_{0}^{(c)}\sqrt{\Nlinks}},
\qquad
\lambda_{c} \in [0,1).
\label{eq:lambdac}
\end{equation}
where $r_{c} = \|\bx_{c(e)} - \bx_{k}\|$ is the current length of chain $c$.
The inverse Langevin function is denoted by $\mathcal{L}^{-1}(\cdot)$, and for compactness, we introduce $\beta_{c}\equiv\mathcal{L}^{-1}(\lambda_{c})$. The chain energy density associated with chain $c$ is then:
\begin{equation}
\psi_{c}^{\text{chain}}
=
\mu\,\Nlinks
\left[
\lambda_{c}\beta_{c}
+
\ln\!\left(
\frac{\beta_{c}}{\sinh \beta_{c}}
\right)
\right],
\label{eq:psichain}
\end{equation}
where $\mu$ is the small-strain shear modulus. The actual energy contribution of chain $c$ is obtained by multiplying Eq.~\eqref{eq:psichain} by its effective reference volume $v_{c}^{\text{eff}}$:
\begin{equation}
E_{c}^{\text{chain}}
=
\psi_{c}^{\text{chain}}\,v_{c}^{\text{eff}}.
\label{eq:Echain}
\end{equation}
where the effective chain volume is $v_{c}^{\text{eff}} = {V_{0,e(c)}}/{8}$, in which $e(c)$ is the cell containing chain $c$.

To match the Abaqus implementation, the inverse Langevin function is approximated using the Cohen approximation~\cite{cohen1991}:
\begin{equation}
\beta_{c}
=
\mathcal{L}^{-1}(\lambda_{c})
\approx
\lambda_{c}
\frac{3-\lambda_{c}^{2}}{1-\lambda_{c}^{2}},
\qquad
\lambda_{c} \le 0.995,
\label{eq:invlangevin}
\end{equation}
which is smooth and monotone for $0 \le \lambda_c < 1$,
has the correct singular behavior as $\lambda_c \to 1^{-}$, and has a maximum relative error of approximately $5\%$ over the admissible range. In the implementation, $\lambda_c$ is clipped at $0.995$ as a numerical safeguard near the locking singularity.
\subsection{Volumetric penalty and near-incompressibility}
\label{sec:volpen}
To enforce the near-incompressibility of elastomers, we used a stored energy with a per-cell volumetric penalty. The energy of cell $e$ is defined as
\begin{equation}
E_{e}^{\text{vol}}
=
\psi_{e}^{\text{vol}}\,V_{0,e}
=
\tfrac{1}{2}K
\left[
\tfrac{1}{2}\left(J_{e}^{2}-1\right)
-
\ln J_{e}
\right]V_{0,e}
\label{eq:psivol}
\end{equation}
where $K$ is the bulk modulus, $V_{0,e}$ is the reference volume of cell $e$,
$J_{e}$ is the cell-averaged Jacobian, and $\psi_{e}^{\text{vol}}$ is strain energy density.

The quadratic-plus-logarithmic form of Eq.~\eqref{eq:psivol} reproduces the
correct small-strain bulk response, vanishes at $J_{e}=1$, and diverges as
$J_{e}\to 0^{+}$. It therefore penalizes strong local compression and provides
a numerical barrier against element inversion. The cell Jacobian is computed as
\begin{equation}
\begin{gathered}
J_{e}
=
\frac{\det \mathbf{A}_{e}}{\det\!\left(\mathbf{B}_{e}+\varepsilon \bI\right)},\\[6pt]
\mathbf{A}_{e}=\sum_{i=1}^{8}\left(\bX_{i}-\bar{\bX}_{e}\right)\otimes\left(\bx_{i}-\bar{\bx}_{e}\right),
\quad
\mathbf{B}_{e}=\sum_{i=1}^{8}\left(\bX_{i}-\bar{\bX}_{e}\right)\otimes\left(\bX_{i}-\bar{\bX}_{e}\right).
\end{gathered}
\label{eq:Jcell}
\end{equation}
where both centroids average over the eight corner nodes,
$\bar{\bX}_{e}=\tfrac{1}{8}\sum_{i}\bX_{i}$ and
$\bar{\bx}_{e}=\tfrac{1}{8}\sum_{i}\bx_{i}$, and $\varepsilon\bI$
regularizes $\mathbf{B}_{e}$ for degenerate cells. $\varepsilon = 10^{-6}$ was used.

\subsection{Total energy}
\label{sec:totalenergy}

The total stored energy of the discrete network is written as the sum of the
chain-network energy and the volumetric penalty energy,
\begin{equation}
E_{\text{tot}}(\bx)
=
E_{\text{chain}}(\bx)
+
E_{\text{vol}}(\bx).
\label{eq:Etot_split}
\end{equation}
The two contributions are
\begin{equation}
E_{\text{chain}}(\bx)
=
\sum_{c\in\mathcal{C}}
\psi_{c}^{\text{chain}}\,v_{c}^{\text{eff}},
\qquad
E_{\text{vol}}(\bx)
=
\sum_{e\in\mathcal{E}}
\psi_{e}^{\text{vol}}\,V_{0,e}.
\label{eq:Etot_components}
\end{equation}

Every chain contribution thus depends on exactly two nodal positions,
and every volumetric contribution on only the eight corner nodes of one cell.
The total energy is therefore a sum of strictly local contributions, which enables vectorized gradient evaluation directly on the GPU (Section~\ref{sec:solution_method}).

\subsection{Homogenization of chain forces into Cauchy stress}
\label{sec:virial}
Although the proposed solution method does not require a continuum stress field during the minimization, we construct a cell-wise Cauchy stress field for comparison with finite-element results by homogenizing chain forces after the equilibrium configuration has been obtained.

For a discrete assembly of central-force interactions, the mean Cauchy stress over a representative volume equals the volume average of the tensor product of the interaction forces with the branch vectors along which they act, a representation established from virtual work by Christoffersen et al.~\cite{christoffersen1981}; specialized to the eight centroid-corner chains of cell $e$ in the deformed configuration, it gives:
\begin{equation}
\bsig_{e}^{\mathrm{chain}}
=
\frac{1}{v_{e}}
\sum_{c \in \mathcal{C}(e)}
f_{c}\,
\frac{\br_{c}\otimes\br_{c}}{r_{c}},
\label{eq:sigma_chain_virial}
\end{equation}
where $\br_{c}$ is the end-to-end vector of chain $c$ in the deformed
configuration, $r_{c}=\|\br_{c}\|$, $v_{e}=V_{0,e}J_{e}$ is the deformed
cell volume, and
\begin{equation}
f_{c}
=
\left\|
\nabla_{\br_{c}}\psi_{c}^{\mathrm{chain}}
\right\|
v_{c}^{\mathrm{eff}}
\label{eq:chain_force_mag}
\end{equation}
For pairwise central forces this branch-vector form coincides with the Irving-Kirkwood virial stress~\cite{irvingkirkwood1950}, whose finite-deformation interpretation was written by Admal and Tadmor~\cite{admal2010}.

The hydrostatic contribution associated with the volumetric penalty is added
separately. Differentiating the volumetric penalty gives
\begin{equation}
p_{e}^{\mathrm{vol}}
=
\frac{1}{2}K
\left(
J_{e}
-
\frac{1}{J_{e}}
\right),
\label{eq:pvol}
\end{equation}
so that the cell-wise Cauchy stress from virial homogenization is:
\begin{equation}
\bsig_{e}
=
\bsig_{e}^{\mathrm{chain}}
+
p_{e}^{\mathrm{vol}}\bI.
\label{eq:sigma_total}
\end{equation}

\section{Solution method: GPU-accelerated L-BFGS}
\label{sec:solution_method}

The proposed method exploits the variational formulation of hyperelastic
 equilibrium and the local structure of the discrete energy in
 Eq.~\eqref{eq:Etot_components}. Each chain term depends on its own end-to-end
 length, and each volumetric term on the volume change of one cell. Their
 gradients can be evaluated in parallel and accumulated into the total energy
 gradient. Minimizing this energy with a matrix-free L-BFGS method avoids
 assembling and factorizing a global tangent stiffness matrix. Figure~\ref{fig:solution}
 emphasizes this contrast; the optimization details are given below.

\begin{figure}[htbp]
\centering
\includegraphics[width=\linewidth]{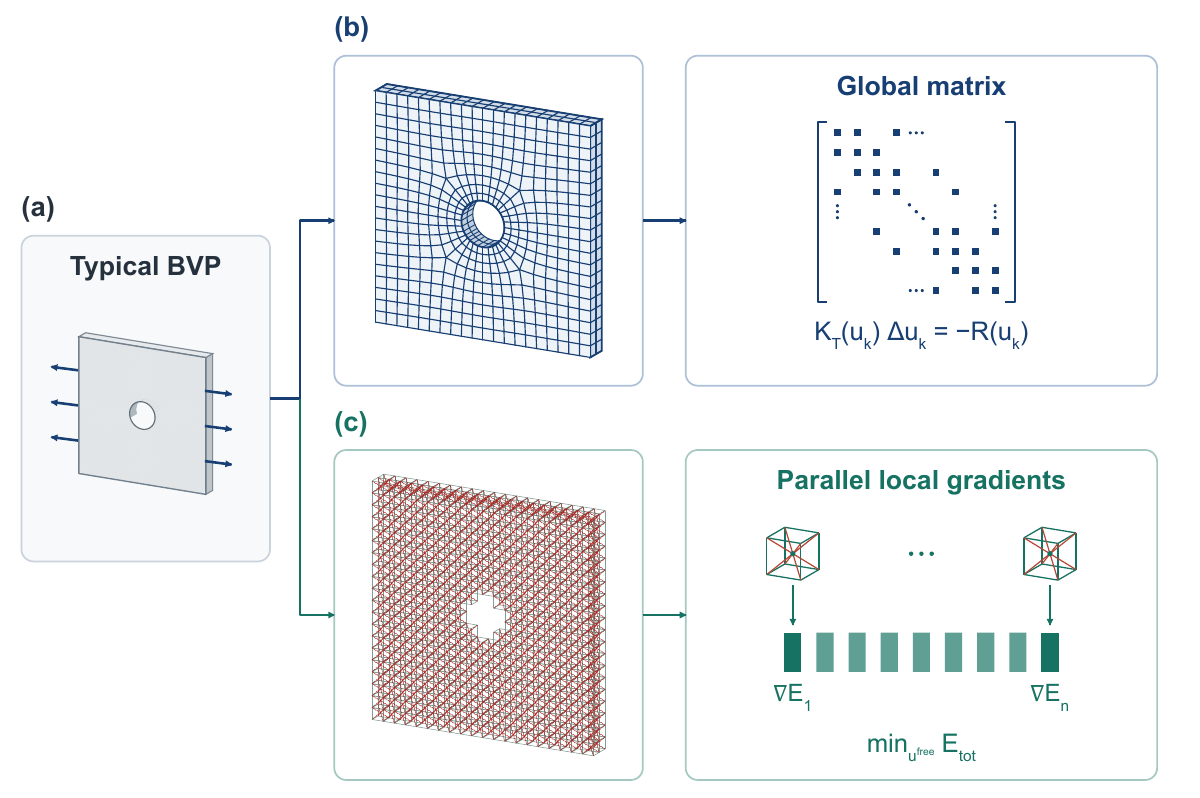}
\caption{
Comparison of conventional finite element analysis and the proposed
matrix-free GPU method for large-deformation hyperelastic boundary-value
problems.
\textbf{(a)} A typical problem: a perforated rubber plate under uniaxial tension.
\textbf{(b)} Conventional FEA discretizes the domain into continuum elements
and assembles a global tangent stiffness matrix.
\textbf{(c)} The proposed method represents the domain using discrete
eight-chain unit cells and minimizes the total energy through parallel
evaluation of local energy gradients. Green rectangles denote cell-local
gradient contributions, which are accumulated at shared nodal degrees
of freedom to obtain the total energy gradient without assembling a
global tangent matrix.
}
\label{fig:solution}
\end{figure}

\subsection{Boundary conditions and parameterization of unknowns}
\label{sec:bc}

The displacement unknowns are parameterized by decomposing the total
displacement field into an affine pre-stretch and a free fluctuation field.
Let $\bu(\bX)$ denote the total displacement field, and let $M$ be a diagonal
zero-one mask whose entries are zero on prescribed displacement components
and one on free components. The displacement field is written as
\begin{equation}
\bu(\bX)
=
\bu^{\mathrm{aff}}(\bX)
+
M\odot \bu^{\mathrm{free}}(\bX),
\label{eq:split}
\end{equation}
where $\bu^{\mathrm{aff}}$ satisfies the imposed macroscopic deformation and
essential boundary conditions, while $\bu^{\mathrm{free}}$ contains the
remaining fluctuation field optimized by L-BFGS. Because the mask removes the
prescribed displacement components from the optimization variables, the
minimization is performed only over the unconstrained degrees of freedom. This
parameterization enforces the Dirichlet constraints exactly throughout the
optimization and reduces the effective search space to mechanically admissible
displacement fields.

For the uniaxial-tension plate benchmarks considered in this work, the specimen
has dimensions $L_X\times L_Y\times L_Z$ and is stretched in the $X$ direction
to a prescribed macroscopic stretch $\lambda$. The displacement boundary
conditions are
\begin{equation}
u_X(-L_X/2) = -(\lambda-1)L_X/2,
\qquad
u_X(L_X/2) = (\lambda-1)L_X/2,
\label{eq:bc_loading}
\end{equation}
together with the symmetry constraints
\begin{equation}
u_X(X=0)=0,
\qquad
u_Y(Y=L_Y/2)=0,
\qquad
u_Z(Z=L_Z/2)=0.
\label{eq:bc_symmetry}
\end{equation}
The affine initialization is chosen as a uniaxial stretch with isochoric
transverse contraction,
\begin{equation}
\lambda_Y=\lambda_Z=\lambda^{-1/2},
\label{eq:isochoric_guess}
\end{equation}
which provides a physically informed initial configuration for the
near-incompressible elastomer. The L-BFGS algorithm then optimizes only the
masked fluctuation field $\bu^{\mathrm{free}}$. Starting from the undeformed
configuration at $\lambda>1$ would require the line search to discover the
macroscopic deformation from scratch and substantially increases the iteration
count.

\subsection{L-BFGS update with zoom line search}
\label{sec:lbfgs}

We minimize $E_{\text{tot}}(\bu^{\text{free}})$ using the limited-memory
BFGS algorithm of Liu and Nocedal~\cite{liu1989}, as implemented in
\texttt{jaxopt}~\cite{jaxopt}. Unless otherwise stated, the L-BFGS history
depth is set to $m=5$, and a zoom line search satisfying Wolfe-type conditions
is used~\cite{nocedalwright2006}. The zoom line search was found to be more robust than simple
backtracking for the present near-incompressible hyperelastic objective, where
the volumetric penalty can produce rapidly changing curvature as $J$ approaches
unity.

Convergence is assessed from the norm of the energy gradient, which
has units of force with respect to the nodal displacement variables. Rather
than prescribing a fixed absolute force tolerance, we define a mesh- and
load-dependent representative chain force from the analytic far-field
eight-chain response under the applied macroscopic stretch. Specifically, for
uniaxial stretch $\lambda$ with isochoric transverse contraction, the
representative chain stretch is
\begin{equation}
\lambda_{\mathrm{ch}}^{\mathrm{ff}}
=
\left(
\frac{\lambda^{2}+2/\lambda}{3}
\right)^{1/2}.
\label{eq:lambda_chain_ff}
\end{equation}
The corresponding inverse-Langevin argument is
\begin{equation}
\eta^{\mathrm{ff}}
=
\frac{\lambda_{\mathrm{ch}}^{\mathrm{ff}}}{\sqrt{\Nlinks}},
\label{eq:eta_ff}
\end{equation}
and the representative volume-weighted spring force is
\begin{equation}
f_{\mathrm{rep}}
=
\left[
\frac{\mu \Nlinks}{\ell_{0}^{\mathrm{ff}}\sqrt{\Nlinks}}
\mathcal{L}^{-1}\!\left(\eta^{\mathrm{ff}}\right)
\right]
v^{\mathrm{eff}},
\label{eq:f_rep}
\end{equation}
where $\ell_{0}^{\mathrm{ff}}$ is the reference body-diagonal chain length of a
cell at the given mesh resolution and $v^{\mathrm{eff}}=V_{0,e}/8$ is the
effective chain volume. The stopping criterion is then
\begin{equation}
\left\|
\nabla_{\bu^{\mathrm{free}}}E_{\mathrm{tot}}
\right\|_{2}
<
\epsilon_{f}\,f_{\mathrm{rep}},
\label{eq:tol}
\end{equation}
where $\|\cdot\|_{2}$ denotes the Euclidean norm over the free displacement
components and $\epsilon_{f}$ is a dimensionless relative force tolerance.
Because $\|\cdot\|_{\infty}\le\|\cdot\|_{2}$, Eq.~\eqref{eq:tol} implies the
pointwise bound
$\|\nabla_{\bu^{\mathrm{free}}}E_{\mathrm{tot}}\|_{\infty}<\epsilon_{f}f_{\mathrm{rep}}$,
so no individual unbalanced nodal force exceeds the prescribed fraction of the
representative far-field chain force. Normalizing by $f_{\mathrm{rep}}$ ties
the tolerance to the material parameters, the applied deformation and the cell
size rather than to an arbitrary absolute value; and because the Euclidean
norm additionally aggregates over all free components, the criterion becomes
more demanding as $\dofs$ increases, so finer meshes are converged at least as
tightly as coarser ones. Unless otherwise stated we use $\epsilon_{f}=0.1$,
chosen from a tolerance-refinement study such that further reduction produces
negligible changes in the reaction force, the local Cauchy stress values, and
the stress-concentration factor.

\subsection{Multiresolution cascade warm start}
\label{sec:cascade}

In addition to the direct solve on the target mesh, the implementation supports
a multiresolution cascade to improve initialization for large problems. Starting
from a coarse structured mesh, the equilibrium displacement field is computed by
L-BFGS and then interpolated trilinearly to the next finer mesh as the initial
guess. This coarse-to-fine procedure is repeated until the target resolution is
reached. The cascade does not change the energy functional or the convergence
criterion; it only provides a physically informed warm start for the final
high-resolution optimization.

When the cascade is enabled, the reported solver wall time includes the
coarse-mesh generation, all L-BFGS iterations across levels, inter-level
trilinear interpolation, and final result output. JIT compilation and the
generation of the target input mesh are reported separately, following the same
timing convention used for comparison with the finite-element benchmark.

\subsection{Locality, parallelism, and the absence of a global tangent}
\label{sec:locality}

Each L-BFGS iteration requires one evaluation of the total energy
$E_{\text{tot}}$ and its gradient with respect to the free displacement
variables. Because every term in Eq.~\eqref{eq:Etot_components} is strictly local, these
quantities can be evaluated through two vectorized local operations on the GPU:
\begin{enumerate}
\item \emph{Chain evaluation.} The $8N_{e}$ chains are evaluated in parallel.
For chain $c$ between nodes $i$ and $j$, the current positions
$\bx_i$ and $\bx_j$ are read, the chain energy
$\psi_{c}^{\text{chain}}$ is evaluated, and its contribution to the energy
gradient with respect to the two endpoint positions is obtained by automatic
differentiation.
\item \emph{Cell evaluation.} The $N_{e}$ cells are evaluated in parallel.
For cell $e$, the eight corner positions are read, the cell Jacobian $J_e$ and
the volumetric energy density $\psi_{e}^{\text{vol}}$ are evaluated, and the
corresponding gradient contribution is computed with respect to the eight
corner nodes.
\end{enumerate}
The chain and cell contributions are auto-vectorized using JAX's
\texttt{vmap} and JIT-compiled into XLA executable graphs. Thus the method
does not assemble a global tangent stiffness matrix and does not require a
linear solve at each iteration. The only global operations are the vector inner
products and vector updates required by the L-BFGS two-loop recursion and line
search.

At iteration $k$, let
\begin{equation}
\mathbf{g}_k
=
\nabla_{\bu^{\mathrm{free}}}
E_{\mathrm{tot}}
\left(
\bu_k^{\mathrm{free}}
\right)
\label{eq:lbfgs_grad}
\end{equation}
denote the gradient of the total energy with respect to the free displacement
variables. L-BFGS constructs a limited-memory approximation to the inverse
Hessian using the most recent $m$ displacement and gradient-difference pairs,
\begin{equation}
\mathbf{s}_k
=
\bu_{k+1}^{\mathrm{free}}
-
\bu_k^{\mathrm{free}},
\qquad
\mathbf{y}_k
=
\mathbf{g}_{k+1}
-
\mathbf{g}_k .
\label{eq:lbfgs_pairs}
\end{equation}
The search direction is then computed as
\begin{equation}
\mathbf{d}_k
=
-\mathbf{H}_k \mathbf{g}_k,
\label{eq:lbfgs_direction}
\end{equation}
where $\mathbf{H}_k$ is the limited-memory approximation to the inverse Hessian
obtained from the stored pairs
$\{(\mathbf{s}_i,\mathbf{y}_i)\}_{i=k-m}^{k-1}$. The free displacement field is updated by
\begin{equation}
\bu_{k+1}^{\mathrm{free}}
=
\bu_k^{\mathrm{free}}
+
\alpha_k \mathbf{d}_k ,
\label{eq:lbfgs_update}
\end{equation}
where the step length $\alpha_k$ is selected by a zoom line search. In the
present implementation, the L-BFGS history depth is $m=5$ unless otherwise
stated.

The full algorithm is summarized in Algorithm~\ref{alg:lbfgs}.

\begin{algorithm}[htbp]
\caption{GPU-accelerated L-BFGS energy minimization for discrete eight-chain elasticity.}
\label{alg:lbfgs}
\begin{algorithmic}[1]
\State \textbf{Input:} reference nodes $\bX$, cell list $\mathcal{E}$, chain list $\mathcal{C}$,
        chain reference lengths $\{\ell_{0}^{(c)}\}$, boundary mask $M$, applied stretch $\lambda$.
\State Compute affine pre-stretch $\bu^{\mathrm{aff}}$ from $\lambda$.
\State Initialize $\bu^{\mathrm{free}} \gets \mathbf{0}$.
\State Initialize the L-BFGS history with depth $m$.
\For{$k = 0,1,2,\ldots$ until converged}
  \State $\bx \gets \bX + \bu^{\mathrm{aff}} + M\odot\bu^{\mathrm{free}}$
  \State $(E,\mathbf{g}_k) \gets \texttt{compute\_energy\_and\_grad}(\bx)$
  \Comment{vectorized chain and cell evaluations}
  \State $\mathbf{d}_k \gets -\mathbf{H}_k\mathbf{g}_k$ using the L-BFGS two-loop recursion.
  \State $\alpha_k \gets \texttt{zoom\_line\_search}(\bu^{\mathrm{free}},\mathbf{d}_k)$
  \State $\tilde{\bu}^{\mathrm{free}}
  \gets
  \bu^{\mathrm{free}} + \alpha_k \mathbf{d}_k$
  \State Compute $\mathbf{s}_k=\tilde{\bu}^{\mathrm{free}}-\bu^{\mathrm{free}}$.
  \State Compute $\mathbf{y}_k=\mathbf{g}_{k+1}-\mathbf{g}_k$ after evaluating the new gradient.
  \State Update the L-BFGS history with $(\mathbf{s}_k,\mathbf{y}_k)$.
  \State $\bu^{\mathrm{free}}\gets\tilde{\bu}^{\mathrm{free}}$.
  \If{$\|\mathbf{g}_k\|_{2}<\epsilon_{f}\,f_{\mathrm{rep}}$}
    \State \textbf{break}
  \EndIf
\EndFor
\State Compute cell-wise Cauchy stress by virial homogenization using Eq.~\eqref{eq:sigma_total}.
\end{algorithmic}
\end{algorithm}

\subsection{Memory footprint}
\label{sec:memory}

In the present implementation, the problem size is reported as
\begin{equation}
\dofs = 3N_{\mathrm{node}},
\label{eq:dof_definition}
\end{equation}
where $N_{\mathrm{node}}$ includes all lattice nodes, including both corner
nodes and cell-center nodes. Dirichlet constraints are enforced by applying the
zero-one mask $M$ to the full displacement vector, rather than by explicitly
eliminating constrained entries. Consequently, the vector storage and L-BFGS
operations scale with the full displacement-vector length $\dofs$.

The memory advantage of the proposed formulation arises from the absence of a
global tangent stiffness matrix. During L-BFGS minimization, the dominant
stored quantities are the displacement vector, the energy gradient, the search
direction, and the limited-memory correction pairs
$\{(\mathbf{s}_{i},\mathbf{y}_{i})\}_{i=k-m}^{k-1}$. Apart from mesh
connectivity and temporary arrays generated by automatic differentiation, the
leading-order vector storage scales as
\begin{equation}
\mathcal{O}\!\left((2m+c)\dofs\right),
\label{eq:memory_lbfgs}
\end{equation}
where $m$ is the L-BFGS history depth and $c$ is an implementation-dependent
constant accounting for auxiliary vectors used by the line search and gradient
evaluation. For fixed $m$, this storage is linear in $\dofs$.

By contrast, an implicit finite-element solver must store the sparse global
tangent matrix and the associated linear-solver data structures. Although the
number of nonzeros in the assembled tangent scales approximately linearly with
$\dofs$ for fixed element order and mesh topology, sparse direct factorization
can introduce substantial fill-in in three-dimensional problems. For regular
3D mesh families, classical nested dissection gives $O(N^{4/3})$ factor
storage and $O(N^2)$ factorization work, where $N$ is the linear-system
dimension~\cite{cambier2020spand}. These are asymptotic reference estimates,
not measured scaling laws for the present thin plate or every Abaqus model;
geometry, ordering, constraints, and pivoting affect practical requirements.
They do not imply quadratic storage of the assembled sparse tangent.
Iterative FE solvers can have different storage and work scaling.
Avoiding global tangent
assembly and factorization therefore reduces memory pressure and helps keep the
large-scale nonlinear solve within the capacity of a single GPU.

\section{The explicit network recovers the Arruda-Boyce response}
\label{sec:validation}

\begin{figure}[!htbp]
\centering
\begin{subfigure}[t]{0.49\linewidth}
    \caption{}\label{fig:val_homogeneous}
    \includegraphics[width=\linewidth]{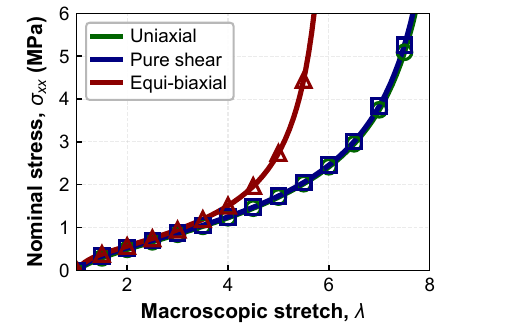}
\end{subfigure}
\hfill
\begin{subfigure}[t]{0.49\linewidth}
    \caption{}\label{fig:val_midplane}
    \includegraphics[width=\linewidth]{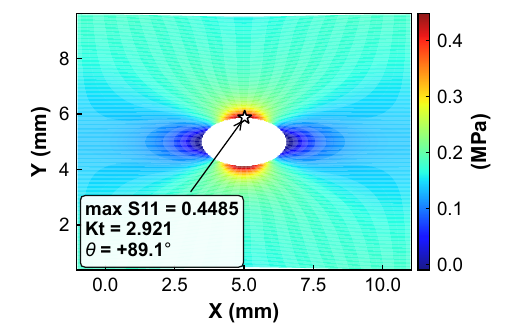}
\end{subfigure}

\vspace{0.8em}

\begin{subfigure}[t]{0.49\linewidth}
    \caption{}\label{fig:val_difference}
    \includegraphics[width=\linewidth]{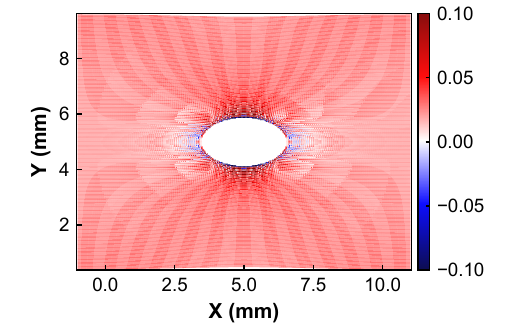}
\end{subfigure}
\hfill
\begin{subfigure}[t]{0.49\linewidth}
    \caption{}\label{fig:val_centerline}
    \includegraphics[width=\linewidth]{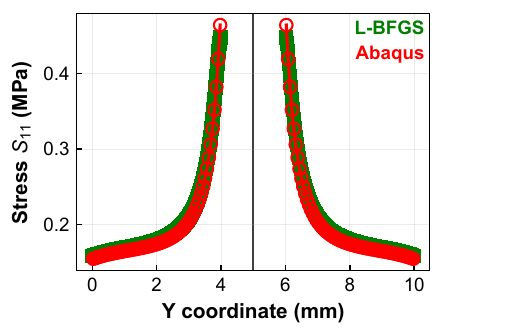}
\end{subfigure}
\caption{
Mechanical validation of the proposed discrete L-BFGS solution method.
\textbf{(a)} Homogeneous stress-stretch response under uniaxial, pure-shear,
and equi-biaxial deformation ($\mu = 0.27$~MPa, $N = 26.5$). Colour denotes the
deformation mode; solid lines are the discrete network response and open
markers the analytic continuum Arruda-Boyce prediction, evaluated with the
exact inverse Langevin function. The uniaxial and pure-shear responses are nearly
coincident on this scale (within $1.6$-$4.6\%$ over the plotted range), with
the pure-shear curve overlying the uniaxial one; the equi-biaxial response
stiffens earliest, as the
larger chain stretch at a given $\lambda$ reaches the limiting extensibility
sooner.
\textbf{(b)} Mid-plane contour of the Cauchy stress component $S_{11}$ for
the plate-with-a-hole benchmark ($320\times320\times32$ lattice,
$1.97\times10^{7}$ degrees of freedom) computed by the proposed method. The
annotated peak ($\max S_{11} = 0.4485$~MPa, $K_t = 2.92$) is the value
recovered by the two-stage post-processing of
Section~\ref{sec:postprocess}; the raw lattice peak is $0.6284$~MPa
($K_t = 4.09$, Fig.~\ref{fig:boundary_smoothing}a).
\textbf{(c)} Normalized signed difference in $S_{11}$ between the proposed
method and Abaqus/Implicit, defined as
$(S_{11}^{\mathrm{L\text{-}BFGS}}-S_{11}^{\mathrm{Abaqus}})/
(F_{\mathrm{react}}/A_{\mathrm{current}})$. The largest differences occur along the stair-stepped hole boundary.
\textbf{(d)} Quantitative comparison of $S_{11}$ along the $Y$-centerline
column adjacent to the hole; the proposed method (green squares, post-processed
per Section~\ref{sec:postprocess}) is compared with Abaqus/Implicit (red
circles). The vertical rule marks the plate centre, $Y = 5.00$~mm; the two
branches are plotted separately and are not bridged across the hole, where this
column carries no material.
}
\label{fig:mechanical_validation}
\end{figure}

Solving the network without the first-invariant closure gives two results
(Fig.~\ref{fig:mechanical_validation}). Under homogeneous deformation aligned
with the cell axes, the eight chains of every cell share the stretch
$\sqrt{I_1/3}$ and the network reproduces the closed-form Arruda-Boyce
stress-stretch response in uniaxial, pure-shear, and equi-biaxial loading.
Under the strongly inhomogeneous deformation of a perforated plate, where the
chains of a cell are free to carry distinct stretches, the network stress field
stays within $3.5\%$ of the continuum Arruda-Boyce finite-element solution over
$95\%$ of the plate (median absolute difference $2.6\%$), and the peak
stress-concentration factor is $3.5\%$ low.

\subsection{Homogeneous deformation: the closure is recovered}
\label{sec:val_homogeneous}

Uniaxial, pure-shear, and equi-biaxial deformations were imposed on a uniform
rectangular specimen with $\mu=0.27$~MPa and $\Nlinks=26.5$, corresponding to a
limiting chain stretch $\lambda_m=\sqrt{\Nlinks}\approx 5.15$. The continuum
reference is the eight-chain Arruda-Boyce relation~\cite{arrudaboyce1993},
\begin{equation}
\sigma_{1}-\sigma_{3}
=
\frac{\mu}{3}
\frac{\sqrt{\Nlinks}}{\lambda_{\mathrm{chain}}}
\mathcal{L}^{-1}
\left(
\frac{\lambda_{\mathrm{chain}}}{\sqrt{\Nlinks}}
\right)
\left(
\lambda_{1}^{2}-\lambda_{3}^{2}
\right),
\label{eq:abcontinuum}
\end{equation}
where $\lambda_{\mathrm{chain}}=\sqrt{\mathrm{tr}\,\bC/3}$. The network response
follows this prediction in all three modes
(Fig.~\ref{fig:mechanical_validation}a). When the loading axes coincide with the
cell axes, each body-diagonal chain attains the same stretch $\sqrt{I_1/3}$, so
the first-invariant closure is not an approximation in this setting: it is the
exact kinematics of the network. The agreement also verifies the chain-energy
implementation and the volume weighting of the chains.

\subsection{Perforated plate: a departure of a few percent}
\label{sec:val_plate}

The perforated plate tests the closure where it is least protected: at finite
strain, with a stress concentration and a curved traction-free boundary. The
specimen has dimensions $L_X=L_Y=10$ and $L_Z=1$, contains a central
through-hole of radius $R=1$, and is stretched in the $X$ direction to
$\lambda=1.2$. The network and the Abaqus/Implicit reference, which solves the
closed-form continuum Arruda-Boyce model, use the same parameters
($\mu=0.27$~MPa, $\Nlinks=26.5$, $K/\mu\simeq100$, the last entered in Abaqus
through the corresponding Arruda-Boyce volumetric parameter) and the same
boundary conditions: the two end faces are prescribed in the loading direction
and symmetry constraints are imposed on the three midplanes.

The network reproduces the full continuum stress field, not only the reaction
force. The mid-plane $S_{11}$ contour (Fig.~\ref{fig:mechanical_validation}b)
shows the tensile lobes above and below the hole and the reduced-$S_{11}$ zones
on its lateral sides. Normalized by the macroscopic stress
$F_{\mathrm{react}}/A_{\mathrm{current}}$, the signed difference from Abaqus
stays within $3.5\%$ over $95\%$ of the plate, with a median absolute
difference of $2.6\%$ (Fig.~\ref{fig:mechanical_validation}c). Along the
$Y$-centerline column adjacent to the hole, the two profiles coincide through
the far field and through the steep rise toward the hole boundary
(Fig.~\ref{fig:mechanical_validation}d). The peak stress-concentration factor
is $K_t=2.921$ for the network and $K_t=3.027$ for Abaqus
(Fig.~\ref{fig:boundary_smoothing}f), a deficit of $3.5\%$.

The largest differences are confined to the immediate vicinity of the hole,
where two effects act together. The first is numerical: the stair-stepped
representation of the circular boundary, treated in
Section~\ref{sec:postprocess}. The second is physical: the freedom of the eight
chains of a cell to depart from their common stretch, which is the content of
relaxing the closure and is isolated on a single cell in
Section~\ref{sec:isotropy}.

\section{Stress recovery at the stair-stepped boundary}
\label{sec:postprocess}

A circular traction-free boundary on a Cartesian grid is a stair-stepped
sequence of retained and removed cells. This approximation has little effect on
the displacement field or on the strain away from the boundary, but cells at
convex turns of the staircase carry locally amplified force per unit deformed
volume, which produces isolated spikes in the recovered Cauchy stress: the raw
lattice peak is $K_t=4.093$ at an off-axis location, against $3.027$ in the
reference. Such artificial concentrations are a known consequence of voxel
representations of smooth interfaces~\cite{fang2016}, and the sensitivity of
local stress estimates to the spatial averaging scale is well documented in
atomistic-to-continuum stress recovery~\cite{hardy1982,zimmerman2004,admal2010}.

\begin{figure}[p]
\centering
\captionsetup{font=footnotesize}
\begin{subfigure}[t]{0.48\linewidth}
    \caption{}\label{fig:bs_raw}
    \includegraphics[width=\linewidth]{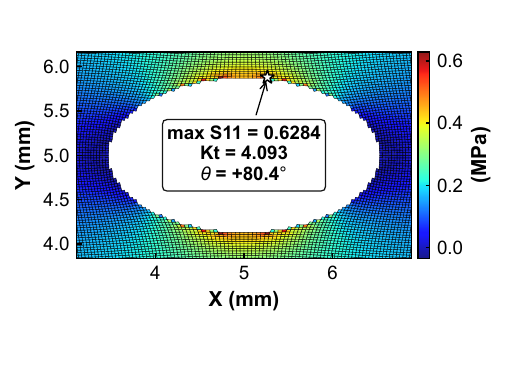}
\end{subfigure}
\hfill
\begin{subfigure}[t]{0.48\linewidth}
    \caption{}\label{fig:bs_target}
    \includegraphics[width=\linewidth]{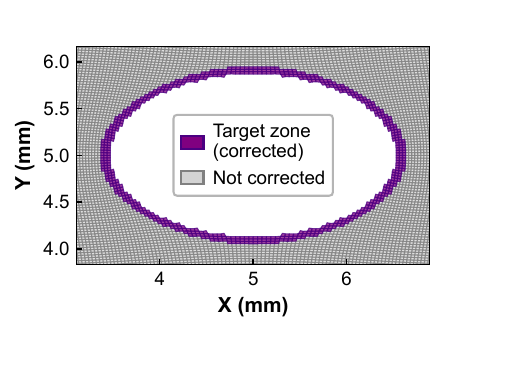}
\end{subfigure}

\vspace{0.4em}

\begin{subfigure}[t]{0.48\linewidth}
    \caption{}\label{fig:bs_stage1}
    \includegraphics[width=\linewidth]{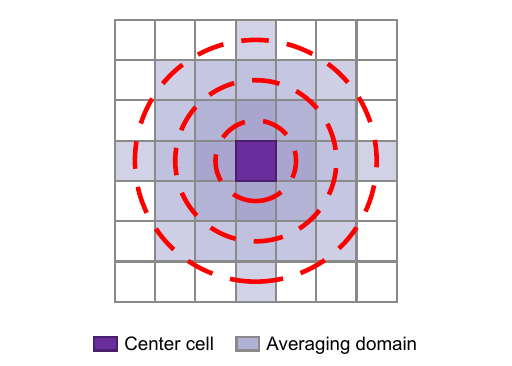}
\end{subfigure}
\hfill
\begin{subfigure}[t]{0.48\linewidth}
    \caption{}\label{fig:bs_stage2}
    \includegraphics[width=\linewidth]{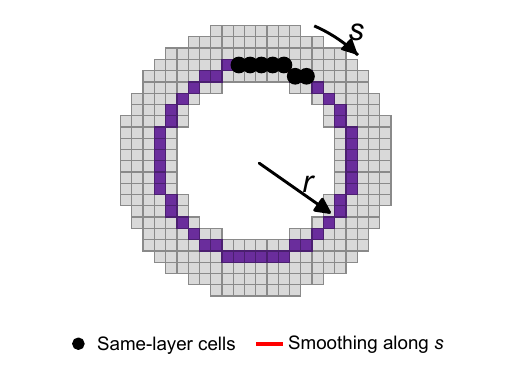}
\end{subfigure}

\vspace{0.4em}

\begin{subfigure}[t]{0.48\linewidth}
    \caption{}\label{fig:bs_recovered}
    \includegraphics[width=\linewidth]{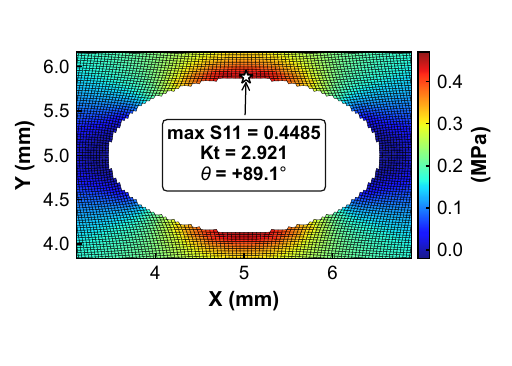}
\end{subfigure}
\hfill
\begin{subfigure}[t]{0.48\linewidth}
    \caption{}\label{fig:bs_reference}
    \includegraphics[width=\linewidth]{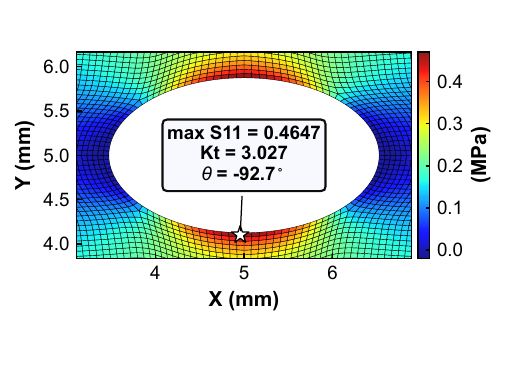}
\end{subfigure}
\caption{
Boundary-aware smoothing of stress spikes at the stair-stepped hole; algorithm
in Section~\ref{sec:postprocess}.
\textbf{(a)} Raw virial \(S_{11}\): an artificial peak of \(0.6284\)~MPa
(\(K_t=4.093\)) at a staircase corner, \(\theta=+80.4^{\circ}\).
\textbf{(b)} Target region: the first three topological layers outward from the
hole.
\textbf{(c)} Stage-1 weighted averaging within \(r_{\mathrm{avg}}=3\Delta x\),
shaded by \(w_{ab}=e^{-d/r_{\mathrm{avg}}}\) (after Ref.~\cite{fang2016}).
\textbf{(d)} Stage-2 smoothing along the boundary tangent, Gaussian width
\(\ell_s=3\Delta x\).
\textbf{(e)} Recovered field: peak \(0.4485\)~MPa (\(K_t=2.921\)) at
\(\theta=+89.1^{\circ}\).
\textbf{(f)} Abaqus/Implicit reference, same colour scale as \textbf{(e)}: peak
\(0.4647\)~MPa (\(K_t=3.027\)), so the recovered peak is \(3.5\%\) low.
}
\label{fig:boundary_smoothing}
\end{figure}

The two-stage recovery below acts on the stress field only. The equilibrium
solution, displacement field, minimized energy, and force balance are
unchanged, and the correction is confined to a narrow zone adjacent to the
hole. With it, the peak returns to the crown of the hole with $K_t=2.921$,
$3.5\%$ below the reference (Fig.~\ref{fig:boundary_smoothing}).

The target zone is defined topologically: layer-1 elements are those directly
adjacent to the removed hole cells, and subsequent layers are grown outward by
in-plane nearest-neighbor expansion. Only the first three layers are corrected;
all other elements retain their original virial stresses.

\paragraph{Stage 1: local weighted spatial averaging.}
For each target element \(a\), the raw stress \(\sigma_{ij}^{(a)}\) is replaced
by a weighted average over a local support domain \(\mathcal{N}(a)\),
\begin{equation}
\tilde{\sigma}_{ij}^{(a)}
=
\frac{
\sum\limits_{b\in\mathcal{N}(a)}
w_{ab}\,\sigma_{ij}^{(b)}
}{
\sum\limits_{b\in\mathcal{N}(a)} w_{ab}
},
\qquad
w_{ab}
=
\exp\!\left(
-\frac{\|\mathbf{x}_{b}-\mathbf{x}_{a}\|}{r_{\mathrm{avg}}}
\right),
\qquad
r_{\mathrm{avg}}=3h ,
\label{eq:fang_discrete}
\end{equation}
where \(\mathbf{x}_{a}\) and \(\mathbf{x}_{b}\) are element-centroid positions
and \(h\) is the in-plane element width (\(\Delta x\) in
Fig.~\ref{fig:boundary_smoothing}). This is the discrete form of the local
stress averaging used for voxel-induced stress
concentrations~\cite{fang2016}. That work averages over one to three square
element layers at a multi-material interface; here the support is Euclidean,
the boundary is traction-free, and the averaging is followed by a tangential
smoothing that has no counterpart in the original method. Stage~1 removes
cell-to-cell speckle while preserving stress variations longer than the
averaging radius.

\paragraph{Stage 2: per-layer arc-length smoothing.}
A residual tangential oscillation remains after Stage~1 because neighboring
boundary cells do not lie on a smooth curve. To remove it without diffusing
stress across radial layers, a one-dimensional Gaussian smoothing is applied
independently within each corrected layer and each \(z\)-slice. The elements of
a layer are ordered along the boundary, assigned a cumulative arc-length
coordinate \(s\), and smoothed with a periodic Gaussian kernel,
\begin{equation}
\hat{S}_{11}(s_a)
=
\frac{
\sum\limits_{b}
\tilde{S}_{11}(s_b)
\exp\!\left[-d_s(s_a,s_b)^2/(2\ell_s^2)\right]
}{
\sum\limits_{b}
\exp\!\left[-d_s(s_a,s_b)^2/(2\ell_s^2)\right]
},
\qquad
\ell_s=3h .
\label{eq:arclength_smoothing}
\end{equation}
where
\begin{equation}
d_s(s_a,s_b)
=
\min\left(
|s_a-s_b|,\,
L_s-|s_a-s_b|
\right)
\label{eq:periodic_arclength_distance}
\end{equation}
is the periodic arc-length distance around a closed layer of total length
\(L_s\). Working in physical arc length keeps the smoothing width at three
element widths along the boundary at every mesh resolution and hole size, and
extends directly to noncircular smooth voids.

\paragraph{Outcome.}
Before correction the maximum stress is a staircase artifact, \(K_t=4.093\) at
\(\theta=+80.4^{\circ}\). After correction the spike is removed and the maximum
sits at the expected crown, \(\theta=+89.1^{\circ}\), with \(K_t=2.921\).
Panels (e) and (f) of Fig.~\ref{fig:boundary_smoothing}, drawn with the same
window, configuration and colour scale, show that the recovered field
reproduces the shape and magnitude of the Abaqus/Implicit reference, with a
residual peak deficit of \(3.5\%\). Under mesh refinement both the magnitude
and the angular location of the recovered \(K_t\) trend toward the reference, which indicates that the residual peak
difference is associated mainly with the finite stair-stepped representation
of the hole.

\section{What the first-invariant closure discards: a residual cubic anisotropy}
\label{sec:isotropy}

\begin{figure}[htbp]
\centering
\includegraphics[width=\linewidth]{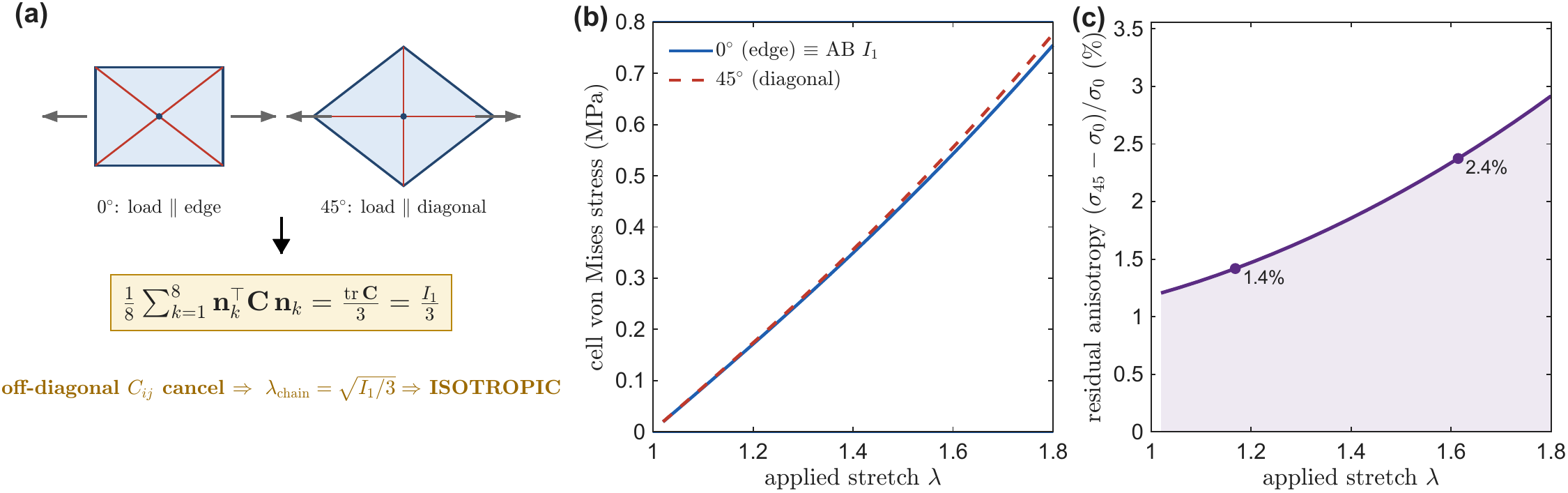}
\caption{Isotropy of the Arruda-Boyce model and the residual anisotropy of the
discrete lattice. \textbf{(a)} The first-invariant collapse: averaging the
squared chain stretch over the eight body diagonals cancels the off-diagonal
components of $\bC$ and leaves only $\mathrm{tr}\,\bC=I_1$, so the chain stretch
$\lambda_{\mathrm{chain}}=\sqrt{I_1/3}$ depends on a single isotropic invariant
and the resulting energy is isotropic. \textbf{(b)} Single-cell uniaxial
von~Mises response for loading along a cell edge ($0^\circ$, which exactly
realizes the $I_1$ collapse) and along a cell face-diagonal ($45^\circ$); the
two are nearly indistinguishable. \textbf{(c)} The residual cubic anisotropy
$(\sigma_{45}-\sigma_{0})/\sigma_{0}$, the kinematic price of evaluating the
chains individually instead of through the $I_1$ collapse: $1.4\%$ at
$\lambda=1.18$, rising to $2.9\%$ at $\lambda=1.8$.}
\label{fig:isotropy}
\end{figure}

The part of the network response that the closure removes can be measured on a
single cell, free of any discretization error. It is a weak cubic anisotropy:
$1.4\%$ at $\lambda=1.18$, rising to $2.9\%$ at $\lambda=1.8$
(Fig.~\ref{fig:isotropy}).

The origin is the averaging step of the Arruda-Boyce construction. A chain
aligned with the unit material direction $\mathbf{n}$ has squared stretch
$\mathbf{n}^{\top}\bC\,\mathbf{n}$, with $\bC=\bF^{\top}\bF$. For the eight
body-diagonal directions $\mathbf{n}_k=(\pm1,\pm1,\pm1)/\sqrt{3}$ the
orientation average is
\begin{equation}
\frac{1}{8}\sum_{k=1}^{8}\mathbf{n}_k^{\top}\bC\,\mathbf{n}_k
=\frac{1}{3}\,\mathrm{tr}\,\bC=\frac{I_1}{3}.
\label{eq:i1collapse}
\end{equation}
The off-diagonal components $C_{ij}$ ($i\neq j$), which carry the orientation
of $\bF$ relative to the cell axes, cancel over the eight sign permutations.
Arruda and Boyce~\cite{arrudaboyce1993} assign this average to every chain,
\begin{equation}
\lambda_{\mathrm{chain}}=\sqrt{I_1/3},
\label{eq:lchain}
\end{equation}
so that the chain energy~\eqref{eq:psichain} depends on the deformation only
through $I_1$ and the strain energy is isotropic: the cubic geometry of the
cell leaves no imprint on the macroscopic response.

The explicit network does not impose
Eqs.~\eqref{eq:i1collapse}-\eqref{eq:lchain}. Each chain is evaluated from its
own end-to-end length, and the virial
homogenization~\eqref{eq:sigma_chain_virial} retains the orientation
information that the average removes. The two descriptions coincide exactly
when the loading axes coincide with the cell axes: for
$\bF=\mathrm{diag}(\lambda,\lambda_{t},\lambda_{t})$ aligned with the cell
edges, all eight chains attain the stretch
$\sqrt{(\lambda^{2}+2\lambda_{t}^{2})/3}=\sqrt{I_1/3}$. When the cell is rotated
relative to the loading, the chains separate into groups with distinct
stretches and the cubic symmetry of the cell reappears in the response.

Figure~\ref{fig:isotropy}(b) compares the uniaxial von~Mises response of a
single cell loaded along a cell edge ($0^\circ$) and along a face diagonal
($45^\circ$), obtained by minimizing the cell energy over the transverse stretch
at each applied $\lambda$. The two curves are nearly coincident. Their relative
difference (Fig.~\ref{fig:isotropy}c) is $1.4\%$ at $\lambda=1.18$, $2.4\%$ at
$\lambda=1.6$, and $2.9\%$ at $\lambda=1.8$; it grows with deformation as the
chains approach locking, where the energy is most sensitive to individual chain
stretches. This anisotropy is intrinsic to the cell and does not vanish under
mesh refinement. It sets the scale of the closure-related departure in a
boundary-value problem, and it is consistent with the few-percent differences
found for the perforated plate in Section~\ref{sec:val_plate}.

\section{Performance and scaling}
\label{sec:performance_scaling}

The solver's cost is governed by local gradient evaluations, and its advantage
over implicit finite-element analysis grows with problem size. Against
Abaqus/Implicit solving the closed-form continuum model on eight CPU threads,
the baseline L-BFGS solver reaches parity near $10^{5}$ degrees of freedom and
a $19.5\times$ speedup at $7.09\times10^{6}$; with the coarse-to-fine cascade
the speedup at that size is $65.3\times$. The lattice was solved up to
$1.97\times10^{7}$ degrees of freedom on a single $24$~GB GPU, with a peak GPU
memory of $15.6$~GB.

\paragraph{Timing and problem-size conventions.}
Abaqus wall times are the wall-clock values in the \texttt{.dat} job summary.
L-BFGS wall times include all L-BFGS iterations, cascade-specific coarse-mesh
generation and interpolation when the cascade is enabled, and final output
writing; they exclude JIT compilation and offline mesh generation, which are
preprocessing. Problem size is reported as
\begin{equation}
\mathrm{DOF}=3N_{\mathrm{node}},
\end{equation}
where $N_{\mathrm{node}}$ is the total number of stored lattice nodes.

\paragraph{Hardware.}
Benchmarks were run on two workstations built around the same processor, an
Intel Core i9-14900KF. The finite-element runs used the machine with $128$~GB of
system memory; the L-BFGS runs used the machine with $64$~GB of system memory
and an NVIDIA GeForce RTX~4090 GPU with $24$~GB of dedicated memory.
Abaqus/Implicit ran on the CPU only and the proposed method on the GPU, so each
solver is reported on the resource that limits it. The CPU is a hybrid design
with eight performance cores at $3.20$~GHz base clock, each carrying two
hardware threads, and sixteen efficiency cores, giving 24 physical cores and 32
hardware threads; $N$ in the parallel-scaling benchmarks denotes hardware
threads. The L-BFGS calculations were executed with JAX/XLA in double precision.

\paragraph{Memory regime of the finite-element baseline.}
Abaqus reports, for each job, the memory required to complete the sparse
factorization without scratch disk traffic: $147$~GB for the
$6.83\times10^{6}$-variable job and $231$~GB for the $9.34\times10^{6}$-variable
job. Both exceed the $128$~GB installed on that machine, so both run partly out
of core, the larger one substantially so; its effective floating-point rate is
about half that of the smaller job on the same machine and thread count. The
speedups reported below are therefore properties of this hardware pairing, not
architecture-independent ratios, and would narrow on a machine able to hold the
factorization in memory.

\subsection{Multicore scaling limits of Abaqus/Implicit}
\label{sec:abaqus_parallel_scaling}

\begin{figure}[!htbp]
\centering
\begin{subfigure}[t]{0.49\linewidth}
    \caption{}\label{fig:scaling_speedup}
    \includegraphics[width=\linewidth]{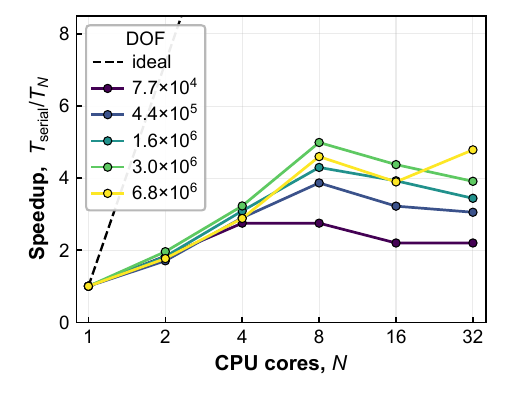}
\end{subfigure}
\hfill
\begin{subfigure}[t]{0.49\linewidth}
    \caption{}\label{fig:scaling_efficiency}
    \includegraphics[width=\linewidth]{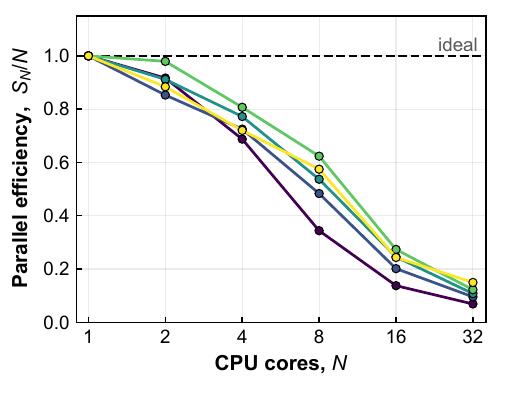}
\end{subfigure}
\caption{
Multicore parallel scaling of Abaqus/Implicit.
\textbf{(a)} Parallel speedup, \(S_N=T_{\mathrm{wall}}(1)/T_{\mathrm{wall}}(N)\),
for five problem sizes. Speedup improves up to four to eight cores and then
saturates or reverses: four of the five series peak at \(N=8\) with a maximum
of \(4.99\times\) and decline thereafter, reflecting communication,
synchronisation and sparse-solver overheads. The \(6.8\times10^{6}\) series is
non-monotonic beyond eight cores, indicating that at the largest sizes the
wall time becomes erratic, not uniformly worse.
\textbf{(b)} Corresponding parallel efficiency \(S_N/N\), for the same five
problem sizes and in the same colours as \textbf{(a)}, which falls
monotonically for every problem size, to \(0.07\)-\(0.15\) at 32 threads. Eight
threads, the full complement of performance cores, is therefore used as the
finite-element baseline throughout.
Abaqus timings are wall-clock values from the \texttt{.dat} job summaries. The \(1.2\times10^{4}\) DOF series is
excluded because its one- to two-second wall times are dominated by timer
quantisation.
}
\label{fig:parallel_scaling}
\end{figure}

Eight threads is the strongest practical configuration of the reference solver
on this hardware (Fig.~\ref{fig:parallel_scaling}). Parallel speedup rises up to
approximately eight threads, with a maximum of $4.99\times$, and then saturates
or declines, while parallel efficiency falls monotonically for every problem
size, to $0.07$-$0.15$ at 32 threads. Part of this is the usual communication,
synchronization, sparse-solver and memory-traffic overhead, but the saturation
point is also architectural: it coincides with the eight performance cores of
the hybrid CPU, so beyond $N=8$ the additional workers are hyperthread siblings
and then efficiency cores, which are slower and cache-poorer. A sparse direct
solve with synchronous phases advances at the rate of its slowest worker, so
adding such workers can reduce throughput outright. The archived CPU-time totals
corroborate this: at $6.8\times10^{6}$ DOF the aggregate CPU time more than
doubles from eight to sixteen threads while the wall time fails to improve.

\FloatBarrier

\subsection{Wall-time scaling and effective speedup}
\label{sec:walltime_speedup}

The wall-time comparison uses Abaqus/Implicit with eight CPU threads pinned to
the performance cores and the UHYPER material, against the RTX~4090 L-BFGS
campaign. Problem size is displacement DOF: three times the number of user
nodes for Abaqus and three times the number of lattice nodes. The hybrid
pressure variables in Abaqus are excluded from this size metric but remain part
of the timed solve. Figure~\ref{fig:parallel_scaling} retains the historical
built-in-material multicore campaign and its total-variable size labels; it is
separate from this UHYPER comparison.

\begin{figure}[!htbp]
    \centering
    \begin{subfigure}[t]{0.49\linewidth}
        \caption{}\label{fig:speed_walltime}
        \includegraphics[width=\linewidth]{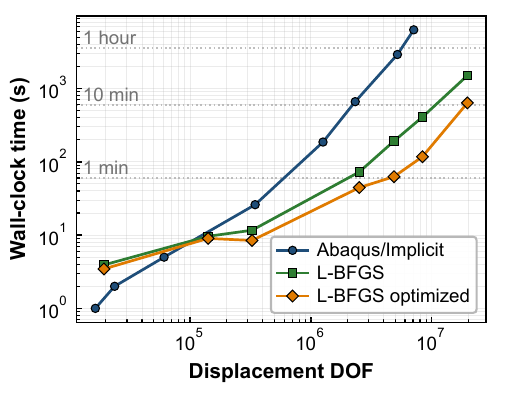}
    \end{subfigure}
    \hfill
    \begin{subfigure}[t]{0.49\linewidth}
        \caption{}\label{fig:speed_effective}
        \includegraphics[width=\linewidth]{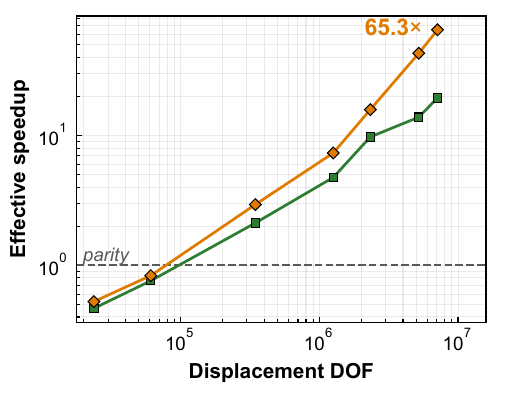}
    \end{subfigure}
    \caption{
    Wall time and effective speedup of the proposed L-BFGS method (RTX 4090)
    relative to Abaqus/Implicit (eight CPU threads pinned to performance
    cores, UHYPER material).
    \textbf{(a)} Wall time versus displacement DOF ($3N$ user or lattice
    nodes). Abaqus is measured from $1.64\times10^{4}$ to $7.09\times10^{6}$
    DOF and the lattice up to $1.97\times10^{7}$ DOF. Below ${\sim}10^{5}$ DOF
    Abaqus/Implicit is slightly faster, because GPU kernel overhead dominates
    and the sparse direct solve is inexpensive; beyond the crossover its wall
    time grows more rapidly.
    \textbf{(b)} Effective speedup
    $T_{\mathrm{Abaqus}}/T_{\mathrm{L\text{-}BFGS}}$, with
    $T_{\mathrm{Abaqus}}$ measured at each Abaqus problem size and
    $T_{\mathrm{L\text{-}BFGS}}$ log-log interpolated onto it from the measured
    lattice solves. Nothing is extrapolated; the smallest Abaqus point lies
    outside the measured lattice range and is omitted. The dashed line marks
    parity. The baseline method crosses parity near $10^{5}$ DOF and reaches
    $19.5\times$ at $7.09\times10^{6}$ DOF; the multiresolution cascade of
    Section~\ref{sec:cascade} reaches $65.3\times$ at the same size, its
    benefit emerging only in the large-DOF regime.
    }
    \label{fig:speed_comparison}
\end{figure}

Figure~\ref{fig:speed_comparison}(a) reports wall time from
$1.64\times10^{4}$ to $7.09\times10^{6}$ DOF for Abaqus and up to
$1.97\times10^{7}$ DOF for the lattice. Below about $10^{5}$ DOF,
Abaqus/Implicit is slightly faster than the baseline L-BFGS solver: each GPU
kernel launch processes too few cells to occupy the device, fixed launch
overhead dominates, and sparse direct factorization is inexpensive. Beyond the
crossover the Abaqus wall time grows more rapidly. In throughput terms,
Abaqus/Implicit falls by more than an order of magnitude across the range, to
$1.12\times10^{3}$~DOF/s at $7.09\times10^{6}$ DOF, while the cascade
implementation sustains ${\approx}2\times10^{4}$~DOF/s over the large-DOF
regime.

Figure~\ref{fig:speed_comparison}(b) reports the effective speedup
$T_{\mathrm{Abaqus}}/T_{\mathrm{L\text{-}BFGS}}$. The two discretizations never
share a degree-of-freedom count, so the ratio is evaluated at the measured
Abaqus problem sizes, with $T_{\mathrm{L\text{-}BFGS}}$ interpolated linearly
in log-log space between measured lattice solves; every Abaqus time is
measured and nothing is extrapolated beyond the measured lattice range. The
baseline solver sits below parity under the crossover and rises steadily
thereafter, reaching $19.5\times$ at $7.09\times10^{6}$ DOF. The cascade of
Section~\ref{sec:cascade} reaches $65.3\times$ at the same size.

The cascade gains only at large DOF, and the mechanism explains why.
Quasi-Newton iteration counts on this class of problems are dominated by slowly
converging long-wavelength modes, whose cost grows with problem size. The
cascade resolves those modes on coarse meshes, so the fine-mesh solve corrects
only the short-wavelength residual; the iterations saved grow with DOF while
the overhead of the coarse solves and inter-level interpolation is essentially
fixed. Accordingly, the iteration-cost reduction grows from $1.9\times$ at
$2.5\times10^{6}$ DOF to $4.8\times$ at $8.4\times10^{6}$ DOF. At small DOF the
coarsest admissible mesh is barely coarser than the target and the fixed
overhead erases the gain.

Two qualifications apply to these figures. First, the steepening of both
speedup curves at the largest sizes is in part a property of the reference
solver's memory regime: the sparse factorization exceeds the installed memory
and the largest job runs furthest out of core. Second, the comparison is made
at equal problem size, not equal accuracy: at $8.4\times10^{6}$ DOF the lattice
recovers a stress-concentration factor $4.7\%$ below the finite-element
reference, so part of its speed advantage is purchased with discretization
error.

\section{Solver robustness on a near-singular mechanism: an auxetic truss metamaterial}
\label{sec:metamaterial}

Because it needs only a gradient, the solver remains well posed where Newton
iteration is not. On a pin-jointed auxetic lattice whose tangent stiffness is
rank deficient, energy minimization returned an admissible configuration at all
ten sizes tested, in $3$ to $128$~ms; Abaqus/Standard succeeded for the two
smallest lattices only, after $8$ to $79$~s per attempt.

The reason is structural. An implicit Newton scheme forms its increment from
$\Delta\bu=-\bK_{\mathrm{T}}^{-1}\br$ and is only as well posed as the tangent
stiffness $\bK_{\mathrm{T}}$; when $\bK_{\mathrm{T}}$ is singular or nearly so,
the increment is undefined. Energy minimization requires only $\nabla E$ and
remains well defined when $\bK_{\mathrm{T}}=\nabla^{2}E$ is rank deficient. A
kinematic mechanism is the extreme case: its tangent stiffness is rank
deficient by the number of zero-energy modes.

\subsection{Benchmark: a re-entrant auxetic honeycomb}
\label{sec:meta_setup}

We use a two-dimensional re-entrant honeycomb lattice, a metamaterial with a
negative Poisson ratio: pulled along one axis, it expands transversely. The
lattice is built from pin-jointed bars connecting vertical struts through
inclined re-entrant links, tiled as an $N_{cx}\times N_{cy}$ array of unit
cells. Each bar is a finite-strain elastic member with the log-strain stored
energy
\begin{equation}
E_{b}=\tfrac{1}{2}\,E A_{0}\,\ell_{0}\,(\ln\lambda_{b})^{2},
\qquad
\lambda_{b}=\frac{\|\bx_{i}-\bx_{j}\|}{\ell_{0}},
\label{eq:bar_energy}
\end{equation}
which is the one-dimensional analogue of the chain energy of
Section~\ref{sec:chainenergy} and coincides with the constitutive response of a
geometrically nonlinear Abaqus \texttt{T2D2} truss element, so that the two
solvers address an identical network. The top boundary is displaced to a $10\%$
vertical stretch with the bottom edge pinned and one node anchored
horizontally, matching the essential-boundary treatment of
Section~\ref{sec:bc}. The proposed method uses the matrix-free L-BFGS minimizer
of Section~\ref{sec:solution_method}; the reference is Abaqus/Standard with
geometric nonlinearity (\texttt{nlgeom}) and automatic incrementation.
Complexity is varied over ten levels, from a single unit cell ($7$ bars) to a
$26\times26$ array ($2107$ bars).

This lattice is a true mechanism. The imposed stretch is accommodated almost
entirely by bar rotation: at the converged states the axial bar stresses are at
the level of numerical round-off ($\max|\sigma|\sim10^{-11}$, against an
$O(10^{2})$ scale for a load-bearing bar), confirming that the deformation lies
in the inextensional null space of the network. The tangent stiffness carries
neither material nor geometric stiffness along these modes and is strongly rank
deficient.

\subsection{Success rate versus complexity}
\label{sec:meta_success}

A solve is counted as \emph{successful} if it converges, at the full applied
stretch, to a configuration free of bar-bar self-intersections. The physical
auxetic response is self-intersection free, whereas a solver that drifts along
the undetermined mechanism modes produces a tangled, self-crossing lattice.
Under this convention the two smallest lattices are counted as finite-element
successes, although Abaqus/Standard converges there to a different branch of
the zero-energy manifold than the energy minimizer returns.

\begin{figure}[htbp]
\centering
\includegraphics[width=1\linewidth]{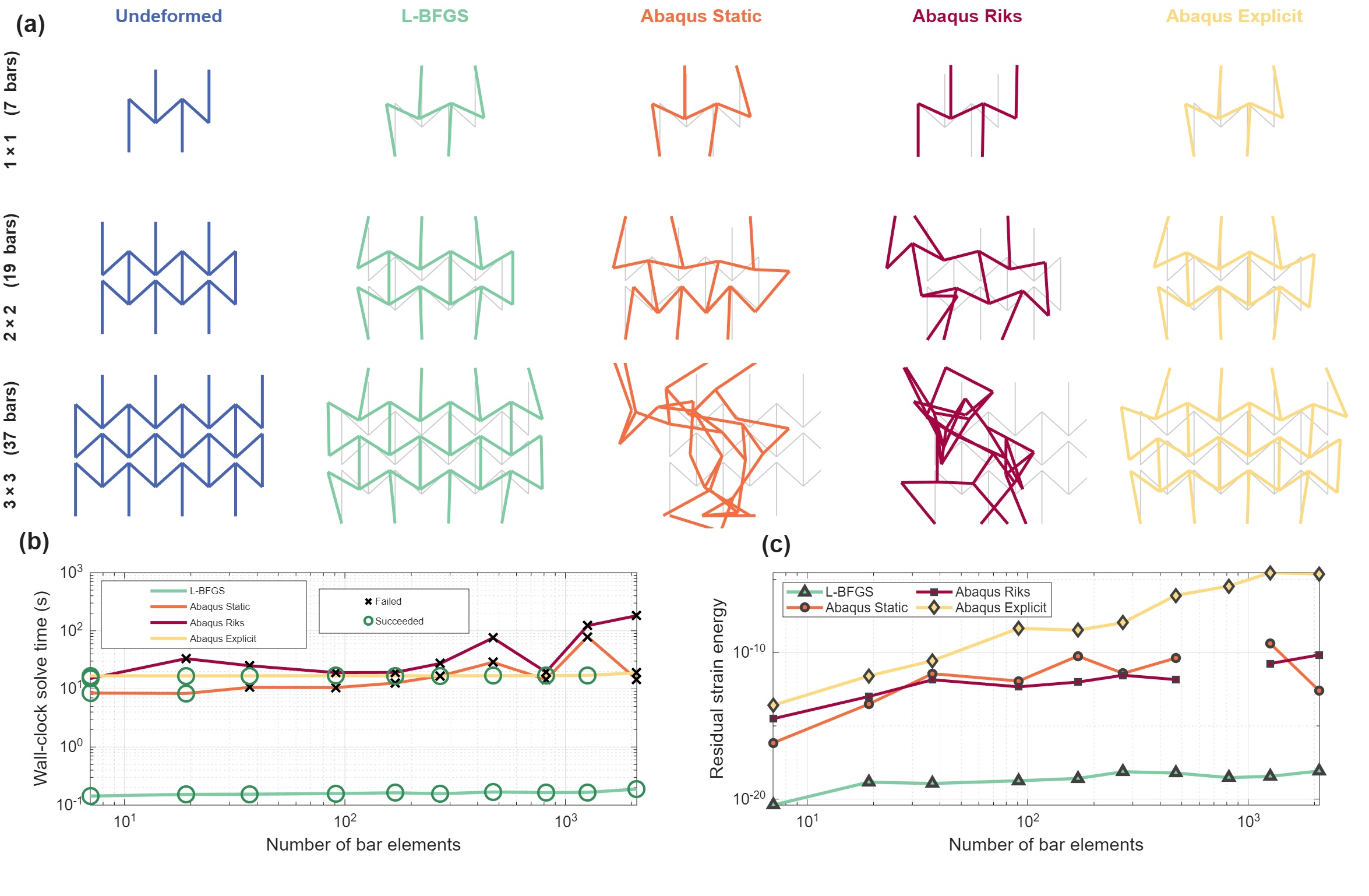}
\caption{
Auxetic truss metamaterial under $10\%$ stretch: deformed configurations and
solver wall time. \textbf{(Left)}~Deformed configurations for the first three
complexity levels ($1\times1$, $2\times2$, $3\times3$ unit cells; $7$, $19$,
$37$ bars). Each row shows, from left to right, the undeformed lattice (blue),
the L-BFGS solution (green), and the Abaqus/Standard solution (red), with the
undeformed lattice underlaid faintly for reference. The L-BFGS solution is
self-intersection free at every size; Abaqus/Standard is clean for the two
smallest lattices but converges to a self-crossing tangle ($24$ crossings) at
$3\times3$. \textbf{(Right)}~Wall-clock solve time versus complexity over the
full sweep (log-log). L-BFGS solves every case in $3$-$128$~ms, whereas
Abaqus/Standard requires $8$-$79$~s and, for all but the two smallest lattices,
expends this time only to fail: green rings mark the two Abaqus successes and
black crosses mark the failures. Over the ten complexity levels the overall
success rate is $100\%$ ($10/10$) for L-BFGS and $20\%$ ($2/10$) for
Abaqus/Standard.
}
\label{fig:metamaterial_success}
\end{figure}

Energy minimization succeeds at every complexity ($10/10$), returning a clean
auxetic configuration with negative Poisson ratio and zero self-intersections,
in solve times from $3$~ms at the smallest size to $128$~ms at $2107$ bars
(Fig.~\ref{fig:metamaterial_success}, Table~\ref{tab:metamaterial}).
Abaqus/Standard succeeds for the two smallest lattices ($2/10$) and fails for
every array larger than $2\times2$ ($19$ bars). It either converges to a
self-intersecting tangle, with the number of crossing bar pairs rising from
$24$ at $37$ bars to $555$ at $469$ bars, or fails to converge at all, in
several cases after substantial effort: the $20\times20$ array consumed $1233$
Newton increments over $79$~s of wall time before diverging, against $69$~ms
for the proposed method on the same problem.

\begin{table}[htbp]
\centering
\caption{Auxetic-metamaterial stretch benchmark: the proposed
energy-minimization solver versus Abaqus/Standard across complexity. Poisson
ratio $\nu$ and solve time are reported for the proposed method; the outcome,
self-crossing count, and end-to-end wall time are reported for
Abaqus/Standard. ``Tangle'' denotes a converged but self-intersecting state;
``diverge'' denotes non-convergence.}
\label{tab:metamaterial}
\small
\begin{tabular}{@{}l r r r r r r r@{}}
\toprule
& & \multicolumn{3}{c}{Proposed (L-BFGS)} & \multicolumn{3}{c}{Abaqus/Standard} \\
\cmidrule(lr){3-5}\cmidrule(lr){6-8}
Cells & Bars & $\nu$ & time & result & outcome & crossings & time \\
\midrule
$1\times1$   & 7    & $-2.20$ & 3\,ms   & \checkmark & success & 0   & 8.5\,s  \\
$2\times2$   & 19   & $-1.64$ & 4\,ms   & \checkmark & success & 0   & 8.3\,s  \\
$3\times3$   & 37   & $-1.49$ & 5\,ms   & \checkmark & tangle  & 24  & 10.6\,s \\
$5\times5$   & 91   & $-1.35$ & 8\,ms   & \checkmark & tangle  & 59  & 10.5\,s \\
$7\times7$   & 169  & $-1.37$ & 19\,ms  & \checkmark & tangle  & 240 & 12.6\,s \\
$9\times9$   & 271  & $-1.37$ & 18\,ms  & \checkmark & diverge & --- & 16.7\,s \\
$12\times12$ & 469  & $-1.37$ & 27\,ms  & \checkmark & tangle  & 555 & 28.7\,s \\
$16\times16$ & 817  & $-1.37$ & 43\,ms  & \checkmark & diverge & --- & 14.5\,s \\
$20\times20$ & 1261 & $-1.36$ & 69\,ms  & \checkmark & diverge & --- & 78.9\,s \\
$26\times26$ & 2107 & $-1.32$ & 128\,ms & \checkmark & diverge & --- & 14.6\,s \\
\bottomrule
\end{tabular}
\end{table}

\subsection{Interpretation}
\label{sec:meta_interpretation}

The contrast follows from the two solution strategies. The mechanism renders
$\bK_{\mathrm{T}}$ rank deficient along the inextensional modes, so the Newton
increment leaves the transverse mechanism degrees of freedom undetermined; the
finite-element solution drifts along the null space and settles on an
arbitrary, load-path-dependent configuration that is typically
self-intersecting, or fails to converge. This is an intrinsic consequence of
requiring the inverse of a singular tangent, not a solver-setting artifact. The
proposed method never forms $\bK_{\mathrm{T}}$. It follows the energy gradient
downhill from the affine warm start of Section~\ref{sec:bc}; along the flat
mechanism directions the gradient is essentially zero, so the minimizer stays
on the ordered branch selected by the initialization.

Two caveats delimit the result. First, a pin-jointed mechanism has a continuum
of zero-energy states, so neither solver defines a unique minimizer; the value
of the proposed method is that it reliably returns a physically admissible
member of that family. Uniqueness would require restoring stiffness to the
mechanism modes, for example through bending resistance in the ribs. Second,
the benchmark demonstrates solver robustness, not constitutive accuracy; it
uses linear-elastic bars and complements the hyperelastic results of
Section~\ref{sec:validation}.

\FloatBarrier

\section{Discussion}
\label{sec:discussion}

\paragraph{The first-invariant closure survives a direct test}
Arruda and Boyce adopted the first-invariant collapse on physical grounds. The
explicit network allows it to be tested directly, and it holds to within a few
percent. Where the
deformation is homogeneous and aligned with the cell, the closure is the exact
kinematics of the network. Where the deformation is strongly inhomogeneous, the
network and the closed-form model differ by $2.6\%$ in the median stress and
$3.5\%$ in the peak stress-concentration factor. These plate figures combine the closure effect with the stair-step
discretization error, which trends downward on refinement, so they do not by
themselves isolate what the closure discards.
The intrinsic part, measured on a single cell with no discretization error, is
a cubic anisotropy of $1.4$ to $2.9\%$ over $\lambda=1.18$ to $1.8$. The
orientation information that the closure averages away therefore carries a few
percent of the stress, concentrated near strong gradients and growing toward
chain locking. The test covers one geometry at a far-field stretch of
$\lambda=1.2$ with the lattice aligned to the loading; lattices rotated
relative to the load and stretches nearer to locking would probe the closure
harder.

\paragraph{Why the solver is fast and robust}
Three coupled ingredients produce both the speed and the robustness: a network
energy that is a sum of strictly local chain and cell terms, a variational
statement of equilibrium, and a quasi-Newton minimizer that needs only
gradients and a short vector history. Together they organize the whole solve
around GPU-resident local evaluations and remove the global tangent, which is
both the dominant cost of implicit analysis at scale and the object whose
singularity defeats Newton iteration on a mechanism. Removing any one
ingredient restores a global operation: a lattice advanced by Newton steps
needs linearized solves, and energy minimization on a conventional
finite-element discretization still requires element quadrature and global
residual assembly.

\paragraph{Limits of the comparison and of the formulation}
The speedups are specific to the hardware pairing and are measured at equal
problem size, not equal accuracy: the reference factorization ran partly out of
core, and the lattice carries a stair-step discretization error at curved
boundaries that a body-fitted finite-element mesh does not. On a mechanism the
minimizer returns an admissible configuration, not a unique one. The
formulation is at present restricted to rate-independent hyperelasticity, to
prescribed displacements and traction-free surfaces, and to regular hexahedral
grids; general tractions, contact, and body-fitted or unstructured cell
arrangements remain to be implemented.

\paragraph{Outlook}
The eight-chain cell was chosen because it is the one network topology whose
closure is known exactly. The solver itself makes no use of that closure, so it
applies unchanged to anisotropic and disordered architectures, to alternative
cell topologies, and to networks with distributed chain lengths, for which no
closed-form closure exists and affine kinematics is least justified. Because
the energy is a sum of local terms, chain scission, damage, and
internal-variable viscoelasticity can be added as further local contributions,
and load continuation follows by warm-starting each increment from the last.

\section{Conclusion}
\label{sec:conclusions}

A matrix-free, GPU-resident L-BFGS minimization of the stored energy makes the
chain network underlying the Arruda-Boyce model directly solvable at finite
strain and at $10^{7}$ degrees of freedom on a single desktop-class GPU. Three
results follow.

First, the first-invariant closure is supported by direct computation. The
network recovers the Arruda-Boyce response under homogeneous deformation
aligned with the cell axes, and under the strongly inhomogeneous loading of a
perforated plate it departs from the closed-form model by a few percent
($2.6\%$ in the median stress, $3.5\%$ in the peak stress-concentration
factor), a difference that includes discretization error. On a single cell,
the closure discards a weak cubic anisotropy of $1.4$ to $2.9\%$.

Second, the cost of the method is governed by local gradient evaluations. The
solver reaches parity with eight-thread Abaqus/Implicit near $10^{5}$ degrees
of freedom and, with a coarse-to-fine cascade, a $65.3\times$ advantage at
$7.09\times10^{6}$, on the hardware tested.

Third, because the method requires only a gradient, it remains well posed on
rank-deficient tangents: it returned an admissible configuration for a
pin-jointed auxetic lattice at all ten sizes tested, where implicit Newton
iteration succeeded at two.

The same solver applies to anisotropic and disordered networks for which no
closed-form closure is available.

\section*{Declaration of generative AI and AI-assisted technologies in the
manuscript preparation process}

During the preparation of this work, the authors used Google Gemini
\cite{gemini} to generate an initial schematic conceptual
illustration for Fig.~\ref{fig:model}, and OpenAI ChatGPT (5.6 Sol)~\cite{chatgpt}
to edit the background of that figure. The final Figure~\ref{fig:model} was
subsequently reconstructed as scripted vector geometry with AI coding
assistance and reviewed and approved by the authors. The authors also used OpenAI ChatGPT~\cite{chatgpt} in preparing Figure~\ref{fig:solution}. Claude was used to refine portions of the text. After using these tools/services, the
authors reviewed and edited the content as needed and take full responsibility
for the content of the published article.

\section*{CRediT authorship contribution statement}

\textbf{Yizhi Chen:} Conceptualization, Methodology, Software, Investigation, Validation, Writing -- original draft.
\textbf{Yuxuan Huang:} Methodology, Software, Investigation, Validation, Writing -- original draft.
\textbf{Zihao Chen:} Software, Data curation.
\textbf{Zhili Xiao:} Software, Data curation.
\textbf{Naiyan Wu:} Investigation, Validation.
\textbf{Dushyanth Rudra Mahesh Velugubantla:} Investigation, Software.
\textbf{Xiangjun Peng:} Methodology, Writing -- review \& editing.
\textbf{Ellen Arruda:} Conceptualization, Writing -- review \& editing.
\textbf{Guy M.\ Genin:} Conceptualization, Supervision, Funding acquisition, Writing -- review \& editing.

\section*{Declaration of competing interest}

The authors declare no competing financial interests or personal
relationships that could have appeared to influence the work reported
in this paper.

\section*{Data and code availability}

The code and data used in this study are available from the authors upon reasonable request.

\section*{Acknowledgements}

The authors acknowledge funding from the NSF through  grants OIA 2219142 and CMMI 1548571.
\bibliography{references}

\end{document}